\documentclass{paper}
\usepackage{fmslash}
\usepackage{mdwlist}

\newcommand{\Lcal}{\mathcal{L}}

\newcommand{\hc}{\text{h.c.}}

\newcommand{\code}[1]{\texttt{#1}}
\newcommand{\url}[1]{\texttt{#1}}

\begin{document}
\begin{titlepage}
  \begin{flushright}
    TTK-10-25
  \end{flushright}        
\vspace{0.01cm}
\begin{center}
{\Large\bf\baselineskip=1.2\normalbaselineskip
  Pseudoscalar Higgs Bosons at the LHC:\\
  Production and Decays into Electroweak Gauge Bosons Revisited\par}
 \vspace{1.5cm}
{\large{\bf Werner Bernreuther\footnote{Email:
{\tt breuther@physik.rwth-aachen.de}},
 Patrick Gonz\'alez\footnote{Email:
{\tt gonzalez@physik.rwth-aachen.de}},
 Martin Wiebusch\footnote{Email: {\tt mwiebusch@physik.rwth-aachen.de}} 
}}
\par\vspace{1cm}
Institut f\"ur Theoretische Physik, RWTH Aachen University, 52056 Aachen,
Germany\\
\par\vspace{1.5cm}
{\bf Abstract}\\
\parbox[t]{\textwidth} {We analyze
   and compute, within a number of standard model (SM) extensions,
  the cross sections $\sigma_{A\to VV'}$ for the production of a
  heavy neutral pseudoscalar Higgs boson/spin-zero resonance at the LHC
   and its  subsequent decays into electroweak gauge
  bosons. For comparison we calculate also the corresponding cross
  sections for a heavy scalar.
   The SM extensions we consider 
    include a type-II two-Higgs doublet model
  (2HDM), a 2HDM with 4 chiral fermion generations, the minimal supersymmetric
  extension of the SM  (MSSM), and  top-color assisted technicolour  models. 
 Presently available phenomenological constraints on  the parameters
 of these models are taken into account. 
    We find that,
  with the exception of the MSSM, these  models permit the LHC cross
  sections $\sigma_{A\to VV'}$ to be of observable size. That is, 
  a  pseudoscalar resonance may be 
  observable, if it exists,  at the LHC  in
  its decays into electroweak gauge bosons, in particular in $WW$ and
  $\gamma \gamma$ final states.}
\end{center}
\vspace*{2cm}
\noindent
PACS number(s):  12.60.-i, 12.60.Fr,  12.60Jv, 12.60.Nz, 14.80.Cp \\
Keywords: Higgs boson decay, weak gauge bosons, standard model 
 extensions 
\end{titlepage}
%
%
%
\section{Introduction}
%
%
The search for Higgs bosons or, more general, (spin-zero) resonances is among
the major physics goals of present-day collider physics, as the existence of
such resonances and the exploration of their properties (production
and decay modes, quantum numbers) would yield decisive clues for unraveling the
mechanism of electroweak gauge symmetry breaking (EWSB).  There is an exhaustive
phenomenology of the production and decay modes of the standard model (SM) Higgs
boson; likewise, there are extensive theoretical studies of these issues for
spin-zero (Higgs) particles predicted by popular SM extensions.  (For reviews
see, e.g., \cite{Djouadi:2005gi} and
\cite{Cvetic:1997eb,Hill:2002ap,Djouadi05,Accomando:2006ga,Morrissey:2009tf}, respectively.)

For the SM Higgs boson $H$ with a mass $m_H \gtrsim 130$ GeV, signatures from
the decay modes\footnote{State-of-the-art predictions for $H\to W W/Z Z \to
  4 \, {\rm fermions}$ were made in \cite{Bredenstein:2006ha}.}  $H\to W W^{(*)}/ Z
Z^{(*)}$ have the highest discovery potential for this particle at the
  Tevatron \cite{Aaltonen:2010h3} and at the Large
Hadron Collider (LHC) \cite{Asai:2004ws,Abdullin:2005yn}.  Concerning
non-standard neutral Higgs particles it is, in view
of unknown model parameters, less clear as to which decay channel
is, for a specific production mode, the most promising one. 
  However,  the decays $A\to W W/ Z
Z$ of a pseudoscalar Higgs boson $A$ are expected to be strongly
suppressed. This is because the couplings $AVV$ $(V=W,Z)$ must be loop-induced,
and they turn out to be very small in two-Higgs doublet extensions (in large
  parts of their parameter spaces)
    and in 
the minimal supersymmetric extension (MSSM) of the SM
\cite{Mendez:1991gp,Gunion:1991cw}. In view of this conventional wisdom one
might be inclined to conclude that the discovery of a spin-zero resonance in
$WW$ and/or $ZZ$ boson events would immediately suggest that it is a
 scalar, i.e., a $J^{PC}=0^{++}$ state. We hasten to add that many suggestions
and phenomenological studies have been made how the spin and the $CP$ parity of
a resonance can actually be measured for these decay modes, irrespective of any
theoretical prejudice \cite{Nelson:1986ki, Soni:1993jc, Skjold:1993jd,
  Barger:1993wt, Arens:1994wd, Choi:2002jk, Buszello:2002uu, Godbole:2007cn,
  Accomando:2006ga, DeRujula:2010ys}.

In this paper we address the question whether there are realistic scenarios
which predict the LHC reactions $p p \to A \to W^+ W^-, \, Z Z$ to be of
observable size. In fact, we analyze a more general class of reactions, namely
the production of a pseudoscalar state $A$ and its decay into electroweak gauge
bosons, $p p \to A \to V V'$, where $VV' \in \{ZZ,WW,\gamma\gamma, Z\gamma\}$.
For comparison we also determine the cross sections $p p \to H\to VV'$ of a
scalar $H$ with mass $m_H \simeq m_A$.  We investigate these cross sections
within several models that contain a CP-odd and two CP-even spin-zero states,
namely the MSSM, a  type II two-Higgs doublet model (2HDM)  and its extension by
a fourth generation of chiral fermions. We briefly address
  also its extension by heavy vector-like quarks.  Due to the
non-decoupling nature of Higgs-fermion couplings, the existence of new heavy
fermions can enhance both the production cross sections of Higgs bosons and the
branching ratios of their decays to $VV'$. We also discuss top-colour assisted
technicolour   (TC2) as a paradigm for scenarios with a relatively light
composite pseudoscalar boson. Within each of these models we determine the largest
possible signal cross sections for $pp\to A\to VV'$ by scanning over the
experimentally allowed region of the respective 
   parameter space. We take
into account the constraints that result from the hadronic branching ratio $R_b$
of $Z\to b {\bar b}$, from flavour observables, 
   from electroweak precision measurements, from direct
Higgs-boson searches at the Tevatron, and from theoretical principles/assumptions.

 For all models discussed below we assume  that the dynamics of the EWSB sector
 is such that the electrically neutral Higgs
resonances are CP eigenstates in the mass basis, at least to very good
approximation. It is well known that Higgs-sector CP violation leads to
neutral spin-zero mass eigenstates that are, in general, a mixture of a CP-odd and
a CP-even component, the latter of which has couplings to $WW/ZZ$ already at
tree-level.

In Section~\ref{sec:approx} we outline the approximations that we used in
  computing the cross sections $\sigma(pp\to A, H\to VV')$, the parameter-space 
  scanning method, the phenomenological constraints, and we list the tools used in this analysis. 
  Section~\ref{sec:THDM} contains our results for the maximal allowed
  cross sections  $\sigma_{A\to VV'}$ and 
  $\sigma_{H\to VV'}$     within a type-II 2HDM with Yukawa couplings
  widely used in the literature. In
  Section~\ref{sec:2HDM4} we extend this analysis to a 2HDM with
   a sequential fourth fermion generation. We comment also on results within
   a 2HDM extended by heavy vector-like quarks. In  Section~\ref{sec:res:MSSM}
      we compute the maximum allowed cross sections 
     $\sigma_{A\to VV'}$  within the 
   so-called phenomenological MSSM (pMSSM) \cite{Djouadi:2002ze}, and in  Section~\ref{sec:TC2}
   the analogous calculations are performed for a composite
   pseudoscalar and a scalar
   spin-zero resonance within TC2. Section~\ref{sec:conclusions}
  contains a summary and our conclusions.

%
%
%
\section{Approximations and Scanning Method}\label{sec:approx}
%
%
 Here we make some general remarks on our
approximations used for the computation of the signal cross sections and the
method we applied for the scans of the respective parameter spaces of
the models below. These models contain two CP-even spin-zero states
$h$ and $H$ (by convention the heavier one is denoted by
 $H$, except in the TC2 models in Sect.~\ref{sec:TC2}) and a CP-odd
 state $A$.    Since we
are mainly interested in the production of $A$ and its decays 
 into massive gauge bosons, we focus on
pseudoscalars  with $m_A\gtrsim \unit{200}{GeV}$. If the mass of $A$ is
significantly above the top-quark pair production threshold then
  $A\to t {\bar t}$ is the dominant  decay mode in
 significant portions of the  parameter spaces of these models. In
 view of our aim of investigating whether or not  the processes 
   $pp\to A \to VV'$
 are relevant for the LHC, we therefore  consider, in the above
 non-SUSY models, a pseudoscalar with $m_A \lesssim 2 m_t$. 
Likewise, the investigation of $pp\to H \to VV'$, made mainly for the
purpose of comparison with the pseudoscalar cross sections, is confined to
scalars $H$ with $m_H \lesssim 2 m_t$. Furthermore, we  compute also the total
Tevatron production cross section for the light Higgs boson $h$ (whose
mass is arbitrary in the non-SUSY models below) and compare with experimental
exclusion limits.

In each model we will use the narrow-width approximation in computing the cross
sections for the production of $\phi=A,H,h$ and
its subsequent decay into $VV'=WW,ZZ,Z\gamma,\gamma\gamma$. To ensure the
validity of this approximation, we constrain model parameters such
    that the total width to mass ratio
$\Gamma_\phi/m_\phi$ is always less than $0.2$.

The dominant Higgs-boson production mechanism at the LHC is gluon fusion.  For all
models discussed below, the corresponding partonic cross sections $\sigma (gg\to
\phi)_{\text{BSM}}$ are calculated in the effective coupling approximation
\cite{Hahn:2006my}. In this approximation the cross section $\sigma(gg
\rightarrow \phi)_{\rm BSM}$ in an SM extension is obtained by rescaling the SM
cross section by the ratio of $\phi\to gg$ decay widths:
\begin{subequations}
\begin{equation}\label{effcouplapprox}
          \sigma( gg \to \phi)_{\text{BSM}}
  \approx \sigma( gg \to H_{\text{ref}})_{\text{SM}}
  \frac{\Gamma(\phi \to gg)_{\text{BSM}}}%
       {\Gamma(H_{\text{ref}} \to gg)_{\text{SM}}}
  \eqpunct,
\end{equation}
where $H_{\text{ref}}$ is a Higgs boson with $m_{H_{\text{ref}}}=m_{\phi}$ and
SM couplings. In some models, for instance  the 2HDM or the MSSM at large
$\tan\beta$, the $b\bar b$ production mode becomes important, too 
\cite{Barnett:1987jw,Dicus:1988cx,Dittmaier:2003ej}. In these
cases we approximate the $b\bar b\to \phi$ cross section analogously by
\begin{equation}
          \sigma(b\bar b \to \phi)_{\text{BSM}}
  \approx \sigma(b\bar b \to H_{\text{ref}})_{\text{SM}}
  \frac{\Gamma(\phi \to b\bar b)_{\text{BSM}}}%
       {\Gamma(H_{\text{ref}} \to b\bar b)_{\text{SM}}}
  \eqpunct.
\end{equation}
\end{subequations}
The SM production cross sections and decay widths were calculated with
\code{FeynHiggs} \cite{Heinemeyer:1998yj}. For the production cross
sections\footnote{See \cite{Harlander:2007zz} for an overwiew of the NNLO QCD computations of
 $\sigma(p {\bar p}, \ pp \to \phi +X)$.}
\code{FeynHiggs} includes NNLO QCD corrections and NNLL soft gluon resummation
effects by interpolating the tables from \cite{Catani:2003zt}. The cross
sections given below refer to the LHC at $\sqrt{s}=\unit{14}{TeV}$.

We have scanned the parameter space of each model, choosing parameter sets
randomly and discarding them if they violate theoretical or experimental
bounds. The theoretical bounds we considered include vacuum stability,
perturbativity and tree level unitarity. On the experimental side we implemented
constraints from direct Higgs boson searches at LEP2 and Tevatron
  by using \code{HiggsBounds} \cite{Bechtle:2008jh}, fits of the
oblique electroweak parameters $S$, $T$, and $U$
\cite{Peskin:1990zt,Altarelli:1990zd}
 and flavour observables measured
in $B$-$\bar B$ mixing and $b\to s\gamma$ decays. More information on these
bounds and their implementation in our analysis is given below
   in the discussions of the
individual models. We used an adaptive sampling technique along the
lines of \cite{Brein:2004kh} in order to 
   find those regions within the allowed parameter space of each model 
 where the signal cross sections are large.

Throughout this paper we will use the following SM parameters:
\begin{gather}
  1/\alpha_{\text{em}} = 137.036
  \quad,\quad
  \alpha_s = 0.118
  \eqpunct,\nonumber\\
  m_Z = \unit{91.19}{GeV}
  \quad,\quad
  m_W = \unit{80.40}{GeV}
  \eqpunct,\nonumber\\
  m_t = \unit{172.6}{GeV}
  \quad,\quad
  m_b = \unit{4.79}{GeV}
  \quad,\quad
  m_\tau = \unit{1.78}{GeV}
  \quad,\quad
  V_{tb} = 1
  \eqpunct.\label{eq:sm_parameters}
\end{gather}
As to the  2HDM
  extensions discussed in this paper, we use
   conventional type-II Yukawa interactions (cf. the comment at the end of
  Section~\ref{sec:THDM}); i.e. 
     the Yukawa couplings of the quarks and leptons of the first and
second generation are assumed to be small. Therefore,
   their interactions with the Higgs resonances will be
neglected in the analysis below.  For the calculation of the decay widths at
one-loop we used \code{FeynArts 3.4} \cite{Hahn:2000kx,Hahn:2001rv} in
combination with \code{FormCalc 6.0} and \code{LoopTools 2.3}
\cite{Hahn:1998yk,Hahn:2006qw}.
%
%
%
\section{Type-II  Two-Higgs Doublet Model}\label{sec:THDM}
%
%
A simple class of SM extensions which contain a pseudoscalar Higgs
particle are the two Higgs doublet models (2HDM), where a second complex Higgs
doublet is added to the SM. Extensive literature exists on these models and
their phenomenological implications \cite{Gunion:1989we}.
   For the convenience of the reader, and in order to fix our notation, we
provide here a brief summary of the model parameters and the physical particle
content.

Following the conventions of \cite{Gunion:2002zf}, we denote the two complex
scalar doublets as $\Phi_1=(\phi_1^+,\phi_1^0)^\trans$ and
$\Phi_2=(\phi_2^+,\phi_2^0)^\trans$. The most general ${\rm SU(2) \times U(1)}$
   invariant tree-level Higgs potential can then be
written as
\begin{align}
  V &=\phantom{{}+{}}
       m_{11}^2\Phi_1^\dagger\Phi_1 + m_{22}^2\Phi_2^\dagger\Phi_2
                                    - [m_{12}^2\Phi_1^\dagger\Phi_2 + \hc]
  \nonumber\\
    &\phantom{{}={}}
      + \tfrac12\lambda_1(\Phi_1^\dagger\Phi_1)^2
      + \tfrac12\lambda_2(\Phi_2^\dagger\Phi_2)^2
      + \lambda_3(\Phi_1^\dagger\Phi_1)(\Phi_2^\dagger\Phi_2)
      + \lambda_4(\Phi_1^\dagger\Phi_2)(\Phi_2^\dagger\Phi_1)
  \nonumber\\
    &\phantom{{}={}}
      + \bigl[  \tfrac12\lambda_5(\Phi_1^\dagger\Phi_2)^2
              + (\lambda_6\Phi_1^\dagger\Phi_1 + \lambda_7\Phi_2^\dagger\Phi_2)
                (\Phi_1^\dagger\Phi_2) + \hc\bigr]
  \eqpunct.
\end{align}
There are strong constraints on $\lambda_6$ and $\lambda_7$, since these are
coefficients of terms which give rise to flavour-changing neutral currents. We
therefore set $\lambda_6=\lambda_7=0$. To ensure $CP$ conservation in the Higgs
sector at tree level we require the remaining parameters to be real without loss
  of generality.  If these parameters are chosen in such a way that
the electric charge is conserved, we can write the vacuum expectation values
(VEVs) of the Higgs doublets as
\begin{equation}
  \langle\Phi_1\rangle = \frac1{\sqrt2}\begin{pmatrix}0\\v_1\end{pmatrix}
  \quad,\quad
  \langle\Phi_2\rangle = \frac1{\sqrt2}\begin{pmatrix}0\\v_2\end{pmatrix}
  \eqpunct,
\end{equation}
with $v_1,v_2\in\R$, and   $v_1^2+v_2^2=v^2\approx(\unit{246}{GeV})^2$ in order
   to  obtain the correct $W$ and $Z$ boson masses.
    After expanding the fields
around their VEVs and diagonalising the mass matrices, the real components of
the Higgs doublets mix and yield two neutral scalar mass eigenstates:
\begin{subequations}
\begin{align}
  H&=(\sqrt2\re\phi_1^0 - v_1)\cos\alpha+(\sqrt2\re\phi_1^0 - v_1)\sin\alpha
  \eqpunct,\\
  h&=-(\sqrt2\re\phi_1^0 - v_1)\sin\alpha+(\sqrt2\re\phi_1^0 - v_1)\cos\alpha
  \eqpunct.
\end{align}
\end{subequations}
By convention, $h$ denotes the lighter of the two states.
  In addition,the physical particle
  spectrum contains one neutral pseudoscalar state $A$ and
   a charged Higgs boson and its conjugate, $H^\pm$.   Expressions for the mixing angle $\alpha$, $v_1$,
$v_2$ and the mass eigenvalues $m_h$, $m_H$, $m_A$, and $m_{H^\pm}$ in terms of
the parameters of the Higgs potential can be found in \cite{Gunion:2002zf}.
Using these expressions, we can describe the 2HDM parameter space by the
following set of independent parameters:
\begin{equation}
  \tan\beta\equiv v_2/v_1\ ,\ \beta-\alpha\ ,\
  m_h\ ,\ m_H\ ,\ m_A\ ,\ m_{H^\pm}\ ,\ \lambda_1
  \eqpunct.
\end{equation}

As already mentioned above, the Yukawa sector we use
    here is that of a  type-II model, i.e.,  the doublet
$\Phi_2$ couples only to up-type fermions and $\Phi_1$ only to down-type
fermions. Experimental bounds on this type of 2HDMs have  been discussed
in several papers, including \cite{Grant:1994ak,Haber:1999zh,Cheung:2003pw,Grimus:2007if}. More
recently a comprehensive study of the allowed parameter space of the ($CP$-violating) type II 2HDM was performed \cite{Kaffas:2007xd}, which combined
several theoretical and experimental constraints. The theoretical constraints
considered in that work are positivity of the Higgs potential, tree-level unitarity
of the $S$ matrix \cite{Ginzburg:2005dt}, and perturbativity. On the
experimental side the relevant bounds come from direct Higgs-boson
   searches at LEP2
and the Tevatron, LEP measurements \cite{Amsler:2008zzb} 
   of the oblique electroweak parameters\footnote{In
     \cite{Froggatt:1991qw,Grimus:2008nb} formulae were 
  given for these parameters for multi-Higgs extensions of the SM.} $S$, $T$
and $U$ and of the ratio $R_b=\Gamma(Z\to
b\bar b)/\Gamma(Z\to\text{hadrons})$. There are also several constraints on the
2HDM parameter space from flavour physics. The strongest ones come from
observables measured in $B$-$\bar B$ mixing and $B\to X_s\gamma$ decays
 \cite{Kaffas:2007xd}.  The
bounds from $R_b$, $B$-$\bar B$ mixing, and $B\to X_s\gamma$ decays only
constrain the parameters $\tan\beta$ and $m_{H^\pm}$ and become relevant for
$m_{H^\pm}\lesssim\unit{300}{GeV}$ or $\tan\beta\lesssim 1$. However, their
combination with bounds from the oblique electroweak parameters can still
constrain the neutral Higgs boson sector, because the states $H^\pm$  with
  a  mass that is very
different from the neutral Higgs-boson masses leads to large contributions to $T$.
Combining the constraints from $R_b$ and the oblique electroweak parameters, we
obtain the lower bound
\begin{equation} \label{2hdmtanb}
\tan\beta \gtrsim 0.7
\end{equation}
 on this parameter. 

Our scans of the parameter space of the 2HDM, which lead in particular to the bound
 (\ref{2hdmtanb}), were made as described in Section~\ref{sec:approx}.
The experimental and theoretical constraints  
  were implemented  by interfacing with
   several publicly available codes.
   Positivity, tree-level unitarity and
perturbativity were checked with \code{2HDMC 1.0.6} \cite{Eriksson:2009ws}. The
perturbativity bound was implemented by requiring that the dimensionless
couplings from the Higgs potential satisfy
$|\lambda_1|,\ldots,|\lambda_5|<4\pi$.  The bounds from direct Higgs-boson
    searches at
LEP2 and the Tevatron were checked with \code{HiggsBounds 1.2.0}
\cite{Bechtle:2008jh}, which provides a model-independent method for deciding 
 whether or not 
a specific parameter is excluded at 95\% C.L.
The oblique
electroweak parameters were calculated with \code{FeynArts 3.4}, \code{FormCalc
  6.0}, and \code{LoopTools 2.3}. The numerical results were compared with those
computed by \code{2HDMC} and perfect agreement was found. In the rele\-vant
parameter space region of the 2HDM the contribution to $U$ is too small to lead to any
constraints. The best-fit values of $S$ and $T$, their standard deviations and
correlation coefficient were taken from \cite{Erler:2010wa,Erler:2010sk}. In our scan
we discarded model parameters that lie outside the 95\% C.L.\ ellipse in the
$S$-$T$-plane. Model parameters that violate the $R_b$ bound at 95\% C.L.\ were
  discarded by using  the respective
   equations from \cite{Haber:1999zh}.  Finally, the flavour
physics bounds are obeyed  by requiring $m_{H^\pm}>\unit{360}{GeV}$.

\begin{figure}
  \centering
  \includegraphics{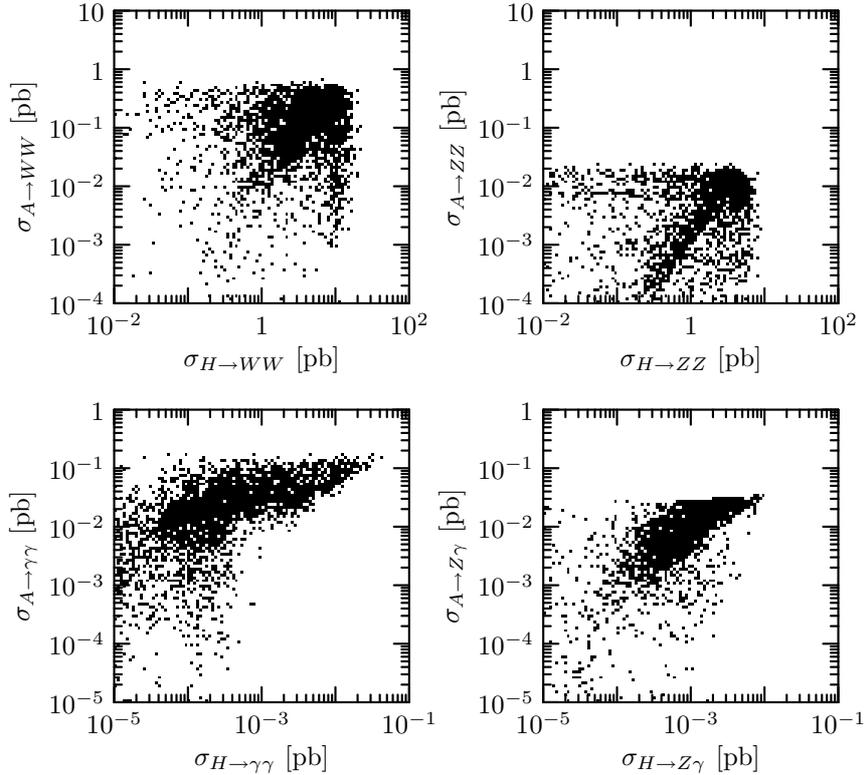}
  \caption{Scatter plots of the LHC cross sections $\sigma_{H\to VV'}$
    versus $\sigma_{A\to VV'}$ in the 2HDM. }
  \label{fig:scatter.thdm}
\end{figure}

With these constraints we have computed the cross sections
$\sigma_{A\to VV'}$,   $\sigma_{H\to VV'}$ for
  the LHC reactions $pp\to A\to VV'$ and $pp\to H\to VV'$ using the approximations
  described in Section~\ref{sec:approx}.
 Figure~\ref{fig:scatter.thdm} shows scatter plots of the combinations
  of $\sigma(pp\to H\to VV')$ and $\sigma(pp\to A\to VV')$
(with $VV'=WW,ZZ,\gamma\gamma,Z\gamma$) that we  found within the allowed
parameter space. Separate parameter scans were performed for each final
state. The importance function (in the sense of \cite{Brein:2004kh}) was set to
$\sigma(pp\to A\to VV')$ if the parameter set passes the constraints discussed
above and to zero otherwise. This means that the scanning algorithm
 seeks out  those regions of the allowed parameter space where 
  $\sigma_A$ is large. The density of the points in 
 Fig.~\ref{fig:scatter.thdm} does therefore \emph{not} represent a probability
density under the assumption of a flat prior. It does, however, give a
qualitative measure of the amount of fine tuning required to obtain certain
combinations of cross sections. The figures show
   that for the $\gamma\gamma$ and $Z\gamma$
final states a large cross section $\sigma_H$ is typically accompanied by a large
  $\sigma_A$. For the $WW$ and $ZZ$ final states there is
no strong correlation. 
  We find that the  maximum values of the  cross sections $\sigma_{A\to VV'}$
    are
\begin{align}
    \sigma(pp\to A\to WW) &\lesssim \unit{0.7}{pb}\eqpunct,
  & \sigma(pp\to A\to ZZ) &\lesssim \unit{0.03}{pb}\eqpunct,
  \nonumber\\
    \sigma(pp\to A\to\gamma\gamma) &\lesssim \unit{0.2}{pb}\eqpunct,
  & \sigma(pp\to A\to Z\gamma) &\lesssim \unit{0.04}{pb}\eqpunct.
\end{align}

There are two scenarios in which the  cross sections $\sigma_{A\to VV'}$ become
maximal simultaneously:
\begin{gather}
  \tan\beta \approx 0.75
  \quad,\quad
  m_A = \unit{320}{GeV}
  \quad,\quad
  m_{H^\pm} > \unit{370}{GeV}
  \eqpunct,\nonumber\\ 
  \beta-\alpha \approx \frac{\pi}{2}
  \quad\text{or}\quad 
  m_h > m_A-m_Z
  \eqpunct,
  \label{eq:thdm:A0par}
\end{gather}
where the values in the first line hold for both scenarios. The 
cross sections $\sigma_A$ are most sensitive to $\tan\beta$ and $m_A$. It is well known
that small values of $\tan\beta$ lead to larger production rates for
pseudoscalar Higgs bosons, because the main Higgs production mechanism at the
  LHC is
$gg\to\phi$ mediated by a top-quark loop with  the $At\bar t$ coupling being
proportional to $\cot\beta$. At the same time the $A\to VV'$ partial widths are
enhanced, because they are dominated by top-quark loops. The above lower bound on
$m_{H^\pm}$ is then necessary for avoiding
   the $R_b$ bound. If  the mass of the pseudoscalar $A$ is sufficiently large but
below the $t\bar t$ threshold then there is no 
  phase-space suppression of the decays
into massive gauge bosons 
   while the competing decay
 channel  $A\to t\bar t$ is closed.
  As  $A\to b\bar b$
  has a small rate for $\tan\beta \sim 1$, the $A\to Zh$ decay would
 be  the dominant one in this case. However, this decay can  
  be parametrically or kinematically
suppressed, which then further increases the $A\to VV'$ branching ratios. This
corresponds to the two options in \eqref{eq:thdm:A0par}. Kinematical suppression
 takes place if the light Higgs boson
   $h$ is sufficiently heavy while  parametric suppression happens if
$\beta-\alpha\approx\pi/2$. This takes us close to the so-called decoupling
limit, which is defined by $m_H,m_A,m_{H^\pm}\gg m_h$ and $\cos(\beta-\alpha)\ll
1$ \cite{Gunion:2002zf}.  In this limit the couplings of 
$h$ to the weak gauge bosons are SM-like, while the (tree-level) couplings 
 of the heavy
Higgs boson $H$ to $WW$ and $ZZ$ and
   thus  $H\to WW,ZZ$ are suppressed. Moreover, the $A\to Zh$
partial width is suppressed by a factor of $\cos^2(\beta-\alpha)$. 
  However, the $pp\to
H\to WW,ZZ$ cross sections can still be of the order of $\unit{10}{pb}$ in this
scenario, if $\beta-\alpha$ is only slightly different from $\pi/2$.

In a second series of scans we took  the  cross sections  $\sigma_{H\to VV'}$ 
to be the respective importance
function and found the following upper limits:
\begin{align}
    \sigma(pp\to H\to WW) &\lesssim \unit{26}{pb}\eqpunct,
  & \sigma(pp\to H\to ZZ) &\lesssim \unit{10}{pb}\eqpunct,
  \nonumber\\
    \sigma(pp\to H\to\gamma\gamma) &\lesssim \unit{0.016}{pb}\eqpunct,
  & \sigma(pp\to H\to Z\gamma) &\lesssim \unit{0.1}{pb}\eqpunct.
 \label{eq:LHCHxsec}
\end{align}
Here the maximal values are also reached for $\tan\beta\approx 0.75$, because
the $Ht\bar t$ coupling is proportional to $1/\sin\beta$; i.e.,  small
$\tan\beta$ increases the $gg \to H$ production cross section.
    Accordingly,  the $R_b$ bound requires the $H^\pm$ to be
sufficiently heavy. Furthermore, $h$ 
should be heavy enough so that $H\to hh$ decays are kinematically
forbidden. However, unlike the above  cross sections for $A$, the  cross
sections  (\ref{eq:LHCHxsec}) do not reach their maximal values 
 in  the same region of parameter space.  The cross
section for the $\gamma\gamma$ final state is maximal for parameters similar to
those in \eqref{eq:thdm:A0par}:
\begin{gather}\label{eq:thdm:H0gamgampar}
  \tan\beta \approx 0.75
  \quad,\quad
  m_H = \unit{265}{GeV}
  \quad,\quad
  m_{H^\pm} > \unit{370}{GeV}
  \eqpunct,\nonumber\\ 
  \beta-\alpha \approx \frac{\pi}{2}
  \quad,\quad 
  m_h > m_H/2
  \eqpunct.
\end{gather}
These values of $\beta-\alpha$ suppress the $H\to WW,ZZ$ decays, which would
otherwise give large contributions to the total $H$ width. Of course, the cross
sections for the $WW$ and $ZZ$ final states are maximal in a region where these
decays are not suppressed:
\begin{gather}\label{eq:thdm:H0WWZZpar}
  \tan\beta \approx 0.75
  \quad,\quad
  m_H = \unit{220}{GeV}
  \quad,\quad
  m_{H^\pm} > \unit{370}{GeV}
  \eqpunct,\nonumber\\ 
  \alpha \approx \frac{\pi}{2}
  \quad,\quad 
  m_h > m_H/2
  \eqpunct.
\end{gather}
The cross section for the $Z\gamma$ final state reaches its maximum in a very
different region of parameter space, namely 
  where the $H\to WW,ZZ$ decays are
kinematically forbidden and the dominant contribution to the $H\to Z\gamma$
decay width comes from bosonic loops:
\begin{gather}\label{eq:thdm:H0Zgampar}
  \tan\beta \approx 0.9
  \quad,\quad
  m_h = m_H = \unit{150}{GeV}
  \eqpunct,\nonumber\\ 
  m_A,m_{H^\pm} > \unit{315}{GeV}
  \quad,\quad
  \alpha \approx \frac{\pi}{2}
  \eqpunct.
\end{gather}

The type II 2HDM we considered here is a popular subject of investigations due
to its close relation to the Higgs sector of the minimal supersymmetric SM. It
is, however, by no means the only possible choice for a 2HDM Yukawa
sector. Extending the present analysis to a wider range of 2HDM models could be
a subject for future studies. An interesting  variant of
a 2HDM, based on the requirement of maximal $CP$ invariance, was proposed in
\cite{Maniatis:2007de}, and its LHC phenomenology was investigated in
\cite{Maniatis:2009vp}. This model contains an SM-like scalar $h_1$, while the
two other physical neutral Higgs boson states $h_2$ and $h_3$ have unusual
  Yukawa couplings.    Moreover, both  $h_2$ and $h_3$
    do not have tree-level couplings to $WW$ and $ZZ$.
%
%
%
\section{A Heavy Fourth Generation}\label{sec:2HDM4}
%
%
Recently, there has been renewed interest in the phenomenology of the SM
extended by a sequential fourth generation of heavy chiral quarks and leptons
(SM4). It was found \cite{Kribs:2007nz,Holdom:2009rf,Hashimoto:2010at} that such fermions with
masses in the (few) hundred GeV range can exist, in spite of strong experimental
constraints.  The presence of a heavy fourth generation would certainly affect
the cross sections for Higgs-boson  production at the Tevatron and
  the LHC, because the dominant
production mode, $gg\to\phi$, would receive additional contributions from loops
of the fourth generation quarks. Additionally, the partial widths of
the loop-mediated $\phi\to VV'$ decays would be affected. Since, in this paper, we are
mainly interested in pseudoscalar Higgs bosons, we will now study the extension
of the type II 2HDM from the last section by a fourth generation of chiral
fermions (2HDM4). We denote the additional fermions by  $u_4$, $d_4$, $\ell_4$
and $\nu_4$ and assume $\nu_4$ to be a Dirac particle.  For simplicity we also
assume that the mixing between the fourth and the first three generations is
strongly suppressed. It is, however, worth noting that  studies
\cite{Alwall:2006bx,Bobrowski:2009ng} of the SM plus a fourth generation with a
general unitary $4\times 4$ CKM matrix showed that large mixing between the
fourth and the first three generations is still allowed experimentally.

The strongest constraints on the masses of the additional fermions come from
direct searches at LEP2 and at the Tevatron, and from electroweak precision
observables. Non-observation at LEP2 implies the lower bounds $m_{\ell_4,\nu_4}
\gtrsim \unit{100}{GeV}$. Searches for heavy quarks at the Tevatron
lead to the bounds
$m_{u_4} > \unit{311}{GeV}$ \cite{Cox:2009mx} and $m_{d_4} >
\unit{338}{GeV}$ \cite{Aaltonen:2009nr}. 
    Experimental bounds on the $S$, $T$,
and $U$ parameters additionally constrain the mass splittings within the $SU(2)$
doublets. Large mass splittings of the $I_W=\pm 1/2$ partners of a doublet yield
large corrections to the $T$ parameter, while small mass splittings result in
large contributions to the parameter  $S$.

For our parameter scans we used the same bounds and tools as for the 
2HDM in Section~\ref{sec:THDM}. We modified our calculations of the $S$, $T$, and $U$
parameters and included the contributions from the extended Higgs
sector and the fourth generation fermions. The latter contributions 
  were compared to the
results from \cite{Kribs:2007nz} and perfect agreement was found. To ensure
perturbativity of the Yukawa sector we required that the Yukawa couplings of the
new fermions lie between $-4\pi$ and $+4\pi$. Finally, the above-mentioned lower
 bounds on the masses of  the fourth generation fermions were used:
\begin{equation}\label{eq:thdm4:mass_bounds}
  m_{\ell_4,\nu_4} > \unit{100}{GeV}
  \quad,\quad
  m_{u_4} > \unit{311}{GeV}  \, , \quad   m_{d_4} > \unit{338}{GeV}
  \eqpunct.
\end{equation}

\begin{figure}
  \centering
  \includegraphics{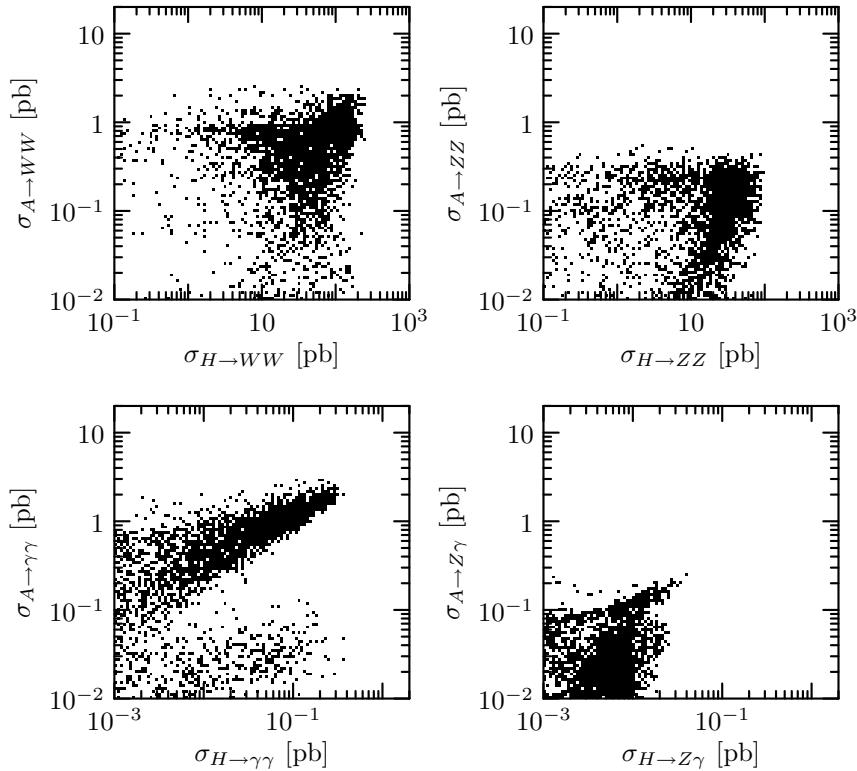}
  \caption{Scatter plots of the LHC cross sections $\sigma_{H\to VV'}$
    versus $\sigma_{A\to VV'}$ in the 2HDM4.}
  \label{fig:scatter.thdm4} 
\end{figure}
Figure~\ref{fig:scatter.thdm4} shows scatter plots of the resulting 
  cross sections $\sigma_{H\to VV'}$ and $\sigma_{A\to VV'}$  
 for the various final states. The maximum
allowed values of the  cross sections for $A$ are:
\begin{align}\label{eq:thdm4:A0limits}
    \sigma(pp\to A\to WW) &\lesssim \unit{3.2}{pb}\eqpunct,
  & \sigma(pp\to A\to ZZ) &\lesssim \unit{0.40}{pb}\eqpunct,
  \nonumber\\
    \sigma(pp\to A\to\gamma\gamma) &\lesssim \unit{3.0}{pb}\eqpunct,
  & \sigma(pp\to A\to Z\gamma) &\lesssim \unit{0.26}{pb}\eqpunct.
\end{align}
They are about one order of magnitude larger than the corresponding bounds  in
the three-generation 2HDM. All the cross sections reach their maximal values in the
following region of parameter space:
\begin{gather}
  \tan\beta \approx 6.3
  \quad,\quad
  m_A \approx \unit{260}{GeV}
  \quad,\quad
m_{H\pm} \approx \unit{360}{GeV}
  \quad,\quad
  \alpha \approx \beta - \frac{\pi}{2}
  \eqpunct,\nonumber\\
  m_{\nu_4}\approx\unit{100}{GeV}
  \quad,\quad
  m_{\ell_4}\approx m_A/2
  \eqpunct.\label{eq:thdm4:A0par}
\end{gather}
The cross sections $\sigma_{A\to VV'}$  have only a weak dependence on the 
  model
parameters that are not stated in (\ref{eq:thdm4:A0par}). Of course, 
  these parameters are subject to  the experimental  and theoretical constraints
 of Section~\ref{sec:approx}.
  Interestingly, a degenerate
fourth generation quark doublet is still allowed in the 2HDM4, even though it
would be excluded in the SM4. The reason is that the large contribution to $S$
from the degenerate quark doublet combined with the large contribution to $T$
from the extended Higgs sector pushes the model back into the 95\% C.L.\ ellipse
in the $S$-$T$-plane. For the parameters above and
$m_{u_4}=m_{d_4}\simeq\unit{330}{GeV}$ the shifts in $S$ and $T$ with respect to the
SM, with a Higgs-boson whose mass is $\unit{117}{GeV}$, are
\begin{equation}
  \Delta S = 0.22
  \quad,\quad
  \Delta T = 0.20
  \eqpunct.
\end{equation}

The main reason for the large  cross sections $\sigma_{A\to VV'}$ 
  in this model is a
dramatic increase of the pseudoscalar production rate $\sigma(gg\to A)$, as
compared to the  2HDM with three generations. 
 This increase is due to additional contributions
from loops of the fourth generation quarks. The couplings of $A$ to up-type (down-type)
quarks are proportional to $\cot\beta$ ($\tan\beta$). 
   As we use conventional type-II Yukawa interactions 
  both types of couplings are proportional to the
 respective fermion mass. Since the 2HDM4 contains a 
  heavy down-type quark $d_4$, the  $Ad_4\bar d_4$ coupling becomes
  strong for large values
of $\tan\beta$. The value of $\tan\beta$ is bounded from above by the
perturbativity constraint on the Yukawa couplings of the fourth generation
fermions, which yields $\tan\beta\lesssim 6.3$. The constraints from 
$R_b$ and $b\to s\gamma$  impose a lower bound on $\tan\beta$ and
   thus on the strength of the $Au_4\bar u_4$ coupling, but it is 
    the maximally allowed value $\tan\beta \approx 6.3$ that leads to the 
  bounds (\ref{eq:thdm4:A0limits}). For this value of $\tan\beta$ the  gluon-fusion
 induced LHC cross section 
$\sigma(gg\to A)$ can become as large as $\sim \unit{840}{pb}$
    in the 2HDM4. The resulting huge 
  pseudoscalar Higgs-boson  production rate 
   would of course dramatically increase
the likelihood for $A$ being discovered
   in  other decay modes like 
$A\to b\bar b$,   $A\to\tau\bar\tau$, or perhaps even in $A\to gg$
dijet events. With the  parameters that yield 
  $\sigma(gg\to A) = 840$ pb we obtain the branching ratios
\begin{equation} \label{eq:2hdm4bgtau}
 B(A\to g g) \simeq \unit{0.46} \, , 
  \quad B(A\to b{\bar b}) \simeq \unit{0.43}   \, , \quad,  
B(A\to \tau^+ \tau^-) \simeq \unit{0.05} \, .
\end{equation}
The $\tau\tau$ final state allows also for a determination of the
 CP quantum numbers of the Higgs-boson resonance \cite{Berge:2008wi,Berge:2008dr}.

If  the values of $\beta-\alpha$ and $m_A$ are as in 
\eqref{eq:thdm4:A0par} and if  $m_H$ is sufficiently large, we are in the decoupling
limit \cite{Gunion:2002zf}. As far as the Yukawa couplings are
concerned, for $\tan\beta \sim 6.3$
    only the couplings of the
heavy Higgs bosons to down-type quarks are enhanced, while the couplings of the
light Higgs boson have  similar magnitudes as the SM
couplings, but the $hu\bar u$ and $h d\bar d$ couplings differ by a sign for
$\alpha>0$. The $u_4$- and $d_4$-loop contributions to the $gg\to h$
amplitude depend sensitively on $\alpha$. For $\alpha$ being small and
positive there are cancellations. In fact, for $\alpha=\pi/2-\beta$
   and $m_{u_4}=m_{d_4}$  the $u_4$- and
   $d_4$-loop would almost exactly cancel, and the $h$ production cross
   section at the Tevatron would essentially be the same as the SM
   cross section. Disregarding this extreme fine tuning we consider a
   set of values $\alpha$ in the vicinity of (\ref{eq:thdm4:A0par})
   and  $m_{u_4}, m_{d_4}$
   such that  the gluon-fusion
 induced  Tevatron production cross section  $\sigma(gg\to
h)$  is about four times
larger than the SM rate. (Without the extended Higgs sector one would expect a
factor of 9.) If one takes into account the
  other production modes that are relevant at the
  Tevatron, $q {\bar q}' \to Wh,\ Zh$, one finds
   that the total Tevatron $h$ production cross section is larger than the SM
    cross section by a factor of about 3.5. This does not invalidate this ``maximum
   scenario''. A closer analysis, using \code{HiggsBounds} \cite{Bechtle:2008jh},
  yields that for the
parameters in \eqref{eq:thdm4:A0par} that a light 
   Higgs boson $h$  with a mass between \unit{145}{GeV}
and \unit{194}{GeV} is excluded by the data from direct Tevatron searches.
  The recently published data  \cite{Aaltonen:2010h3}  will widen this exclusion window,
  but  will not falsify the above scenario.

While pseudoscalar Higgs-boson production via gluon fusion can only be mediated by quark loops, the
decays into electroweak gauge bosons can also be mediated by loops of leptons or
neutrinos. As the mass bounds from direct searches for fourth generation leptons
are still relatively low, it is possible to construct a situation where $m_A$ is
close to the $\ell_4\bar\ell_4$ threshold. In that case the $\ell_4$-loop
contribution to $\Gamma(A\to VV')$ is kinematically enhanced, which leads to
larger branching ratios. However, since fixed order calculations are unreliable
at threshold we conservatively require a minimum difference of \unit{10}{GeV}
between $m_{\ell_4}$ and $m_A/2$. The upper limits \eqref{eq:thdm4:A0limits} for
the pseudoscalar cross sections were obtained with this requirement and might
actually be slightly bigger. For the $ZZ$, $\gamma\gamma$ and $Z\gamma$ final
states the branching ratios could be further increased by setting
$m_{\nu_4}\approx m_A/2$, too. However, it turns out that the constraints on $S$
and $T$ forbid such a choice of parameters.

The upper limits on the cross sections for the heavy scalar Higgs boson are:
\begin{align}\label{eq:thdm4:H0limits}
    \sigma(pp\to H\to WW) &\lesssim \unit{280}{pb}\eqpunct,
  & \sigma(pp\to H\to ZZ) &\lesssim \unit{110}{pb}\eqpunct,
  \nonumber\\
    \sigma(pp\to H\to\gamma\gamma) &\lesssim \unit{0.41}{pb}\eqpunct,
  & \sigma(pp\to H\to Z\gamma) &\lesssim \unit{0.045}{pb}\eqpunct.
\end{align}
The cross sections associated with the
loop-induced $H$ decays, $\sigma(pp\to H\to \gamma\gamma,Z\gamma)$,
    become maximal simultaneously with the  cross
sections $\sigma_{A\to VV'}$, i.e.\ for the parameters \eqref{eq:thdm4:A0par}. For the cross
sections $\sigma(pp\to H\to WW,ZZ)$ the relevant parameters are $\tan\beta$,
$m_H$, $\beta-\alpha$ and $m_h$. The maximal values are reached for
\begin{equation}
  \tan\beta \approx 5.7
  \quad,\quad
  m_H \approx \unit{210}{GeV}
  \quad,\quad
  \beta-\alpha \approx \frac\pi4
  \quad,\quad
  m_h > m_H/2
  \eqpunct.
\end{equation}
This can be understood as follows: the couplings of $H$ to up-type quarks are
proportional to $\sin\alpha/\sin\beta$ while its couplings to down-type quarks
are proportional $\cos\alpha/\cos\beta$. Therefore, for reasons analogous to those
discussed for the $A$ production cross sections, the largest $H$ production rates
are obtained for large $\tan\beta$, i.e.\ for an enhanced $Hd_4\bar d_4$
coupling. This coupling is increased further if the mixing angle $\alpha$ is small. 
 However,
the (tree-level) partial widths for $H\to WW,ZZ$ are proportional to
$\cos^2(\beta-\alpha)$ and would be suppressed for large $\tan\beta$ and a small
$\alpha$. In the search for the largest cross
  sections $\sigma(gg\to H\to WW,ZZ)$, the best compromise
turns out to be the choice $\beta-\alpha=\pi/4$. Within the region  of
    $m_H$ values
   allowed by direct
Higgs-boson  searches, the $gg\to H\to WW,ZZ$ cross sections steadily increase for
decreasing $m_H$, i.e.,  the largest cross sections are obtained for relatively
small $m_H$. Finally the mass $m_h$ of the light Higgs boson 
  must be large enough so that the competing $H\to hh$
decay mode is kinematically forbidden. The masses of the fourth generation
fermions have very little influence on the $gg\to H\to WW,ZZ$ cross sections in
this scenario, as long as they are  in agreement with the mass
bounds \eqref{eq:thdm4:mass_bounds} and the constraints on $S$ and $T$.

A more exotic possibility for new heavy fermions are vector-like quarks, i.e.,
quarks whose left- and right-chiral components have equal gauge charges (see,
e.g. \cite{Frampton:1999xi,Aguila00,delAguila:2000aa,AguilarSaavedra:2009es}).
Such states are present in a number of SM extensions, including extra
dimensional models with bulk fermions \cite{Appelquist:2000nn} and Little Higgs
models \cite{ArkaniHamed:2002qy}. In \cite{Bernreuther:2009ts} we presented an
extension of the 2HDM by one ${\rm SU(2)}$ singlet and two doublets of vector-like quarks and
computed the loop contributions of these new fermions to the pseudoscalar decay
widths. However, if the constraints from the oblique electroweak parameters are
applied to this model it turns out that these contributions do not alter the
results presented for the three-generation 2HDM in section~\ref{sec:THDM} in any
noticeable way.
%
%
%
\section{The MSSM}\label{sec:res:MSSM}
%
%
The decays of the scalar Higgs bosons in the MSSM have already been discussed in
detail in the literature. (See \cite{Djouadi05} for a review.) In this section
we will therefore concentrate on the cross sections $\sigma(pp\to A\to VV')$ for
the pseudoscalar Higgs boson $A$. In \cite{Gunion:1991cw} the branching ratios
of $A$ into gauge bosons were calculated under the assumption that all SUSY
particles are too heavy to yield relevant contributions to the effective $AVV'$
couplings. However, experimental bounds on chargino and neutralino masses are
still relatively weak and values as low as \unit{100}{GeV} or \unit{60}{GeV},
respectively, are still possible.\footnote{If the requirement of gauge
  unification at the GUT scale is dropped, the lightest neutralino could even be
  massless.} We have therefore extended the results of \cite{Gunion:1991cw} to
include the contributions of SUSY particles to the loop-mediated
$gg\to A$
production mechanism and the $A \to VV'$ decays. We will see that these
contributions are relevant in those regions of the MSSM parameter space that
maximize the $pp\to A\to VV'$ cross sections. To avoid misunderstandings: we consider
 here the MSSM with three generations.

The Higgs sector of the MSSM is that of a type-II 2HDM discussed in section
\ref{sec:THDM}, but with a much more restricted parameter
  space.  In the MSSM the masses of the neutral Higgs bosons and the
mixing angle $\alpha$ are no longer independent parameters. At tree level they
can be expressed in terms of the pseudoscalar Higgs-boson mass $m_{A}$ and
$\tan\beta$ as follows:
\begin{subequations}
\label{eq:tree_level_relations}
\begin{gather}
  m_{H^\pm}^2 = m_A^2 - m_W^2
  \eqpunct,\label{eq:m_A}\\
  m_{H,h}^2 = \frac12\left[m_A^2+m_Z^2\pm
  \sqrt{(m_A^2+m_Z^2)^2 - 4m_Z^2m_A^2\cos^22\beta}\right]
  \eqpunct,\\
  \cos(2\alpha) = -\cos(2\beta)\frac{m_A^2-m_Z^2}{m_H^2-m_h^2}
  \quad,\quad
  \sin(2\alpha) = -\sin(2\beta)\frac{m_H^2+m_h^2}{m_H^2-m_h^2}
  \eqpunct.
\end{gather}
\end{subequations}
For MSSM scenarios with $m_A\gg m_Z$ these equations yield
\begin{equation}\label{eq:approximations}
  m_A\approx m_H \quad,\quad \beta-\alpha\approx\frac\pi2
  \eqpunct.
\end{equation}
However, it is well-known that the tree-level relations
\eqref{eq:tree_level_relations} are substantially modified by loop corrections
to the MSSM Higgs potential. These corrections are responsible for pushing 
the mass of the light Higgs boson substantially above the $Z$-boson mass and have
to be taken into account to obtain reliable results. In our scans we used
\code{FeynHiggs 2.6.5} \cite{Frank:2006yh, Degrassi:2002fi, Heinemeyer:1998np,
  Heinemeyer:1998yj} to calculate all one-loop and leading two-loop corrections
to the neutral Higgs-boson self-energies in the MSSM and to extract from them the
physical neutral Higgs-boson masses, LSZ residues, and the resulting effective mixing
angle $\alpha_{\text{eff}}$. We also use \code{FeynHiggs} for the calculation of
the total Higgs-boson decay widths.

As mentioned above, contributions from SUSY particles have to be taken into
account when calculating the amplitudes for the loop-mediated production or
decay processes. At the one-loop level, the decay amplitudes of the Higgs bosons
 receive also contributions from loops of squarks, sleptons, charginos and
neutralinos. However, the squark and slepton loop contributions to the
pseudoscalar Higgs decays vanish since parity is conserved in the bosonic sector
of the MSSM. For the same reason the $gg\to A$ production amplitude receives no
new contributions, while the $gg\to H$ process is now also mediated by
squark loops.

The MSSM contains a large number of parameters due to the 
 soft SUSY-breaking part of the Lagrangian. To make phenomenological studies
feasible, the number of free parameters has to be reduced significantly. In this
paper we will work in the so-called phenomenological MSSM (pMSSM)
\cite{Djouadi:2002ze}, where the number of parameters is reduced to 22 by making
several phenomenologically motivated assumptions, including the absence of both
new $CP$-violating phases and new sources of flavour violation. The independent
parameters of the pMSSM are
\begin{itemize}
\item $\tan\beta$ and $m_A$,
\item the Higgs-Higgsino mass parameter $\mu$,
\item the gaugino mass parameters $M_1$, $M_2$, and $M_3$,
\item the light sfermion mass parameters $m_{\tilde q}$, $m_{\tilde u_R}$,
  $m_{\tilde d_R}$, $m_{\tilde l}$, and $m_{\tilde e_R}$,
\item the light sfermion trilinear couplings $A_u$, $A_d$, and $A_e$,
\item the third generation sfermion mass parameters $m_{\tilde Q}$, $m_{\tilde
  t_R}$, $m_{\tilde b_R}$, $m_{\tilde L}$, and $m_{\tilde\tau_R}$,
\item the third generation trilinear couplings $A_t$, $A_b$, and $A_\tau$.
\end{itemize}
For the exact definition of these parameters and a discussion of the assumptions
under which the MSSM parameter space reduces to this subset, we refer the reader to
\cite{Djouadi:2002ze}. Scanning this 22-dimensional parameter space and
implementing the relevant experimental bounds, in particular those from direct
searches of SUSY particles at the Tevatron, is still a difficult task
\cite{Berger:2008cq}. Fortunately the cross sections that we study in this paper
turn out to be insensitive to most of the parameters above. As a result we still
obtain reliable upper limits for the cross sections by scanning over an even
smaller number of para\-meters and imposing conservative sparticle mass limits in
order to satisfy the Tevatron bounds. Let us therefore take a moment to motivate
our choice of independent variables for the parameter scans.

In the three-generation 2HDM we obtained the largest (scalar and pseudoscalar) Higgs production
cross sections for $\tan\beta\lesssim 1$, because the $gg\to\phi$ production mechanism is
enhanced in that case. In the pMSSM the bounds on the lightest Higgs-boson mass (which
now depends on $\tan\beta$) require $\tan\beta\gtrsim 3$. In that case the Higgs
production rate due to gluon fusion is much smaller than for
$\tan\beta\sim 1$. However, for very large values of $\tan\beta$ the $\phi b\bar b$
couplings are enhanced and the $b\bar b\to\phi$ production mechanism (at the
LHC) can become the dominant one. In the pMSSM the largest Higgs-boson production
cross sections are obtained in this scenario. Of course the $\phi\to b\bar b$
partial decay widths then dominate the total width and the branching ratios for
other decay modes are suppressed. Nonetheless, the largest $pp\to\phi\to VV'$
cross sections are obtained at large $\tan\beta$. In this case the production
cross sections $\sigma(pp\to\phi)$ and the total decay widths $\Gamma_\phi$ are
both approximately proportional to the strongly enhanced $\phi b\bar b$
coupling. Thus any dependence on the pMSSM parameters that enters through loop
corrections to the $\phi b\bar b$ vertices cancels when we compute the signal
cross sections $\sigma(pp\to\phi\to VV')$, because they are proportional to the
ratio $\sigma(pp\to\phi)/\Gamma_\phi$.

For determining the relevant pMSSM parameters it is therefore sufficient to look at
the partial widths of the $\phi\to VV'$ decay processes.  As explained earlier,
the only SUSY particles contributing to the $A\to VV'$ amplitudes are charginos
and neutralinos. Their masses and couplings depend only on the parameters
$\tan\beta$, $\mu$, $M_1$, $M_2$, and $m_A$. The latter parameter 
  enters through the Higgs
mixing angle $\alpha$.  A dependence on the other parameters is introduced if
loop corrections to $\alpha$ are taken into account. The loop corrections to
$\alpha$ are calculated from self-energy corrections to the neutral scalar Higgs
propagators. The same self-energies determine the physical mass of the light
Higgs boson, which in turn is subject to strong experimental constraints. The
dominant contributions to these self-energies come from loops of top quarks and
top squarks, due to the large top Yukawa coupling.  We therefore expect our
results to be also sensitive to those pMSSM parameters which affect the top
squark masses and couplings, i.e.\ $m_{\tilde Q}$, $m_{\tilde t_R}$, and
$A_t$. For the $H\to WW,ZZ$ decays the situation is even simpler: the scalar
Higgs bosons couple to $W$ and $Z$ bosons at tree level with couplings
proportional to $\cos(\beta-\alpha)$, and the dominant SUSY corrections to these
vertices are obtained by replacing $\alpha$ by $\alpha_{\text{eff}}$ in the
vertex factors.

These considerations motivate us to set
\begin{gather}
  A_u = A_d = A_e = A_\tau = A_b = 0
  \eqpunct,\nonumber\\
  m_{\tilde q} = m_{\tilde u_R} = m_{\tilde d_R} = m_{\tilde l} = m_{\tilde e_R}
  = m_{\tilde Q} = m_{\tilde b_R} = m_{\tilde L} = m_{\tilde\tau_R}\equiv m_S
  \eqpunct.
\end{gather}
Furthermore we impose the GUT relation
\begin{equation}
  M_1 = \frac53\tan^2\theta_W\,M_2
\end{equation}
and use
\begin{equation}\label{eq:mssm:parameters}
  \tan\beta\ ,\ m_A\ ,\ \mu\ ,\ M_2\ ,\ M_3\ ,\ m_{\tilde t_R}\ ,\ A_t\ ,\ m_S
\end{equation}
as independent variables for the parameter scans. On this reduced parameter
space we apply the experimental constraints from direct Higgs-boson searches, as
explained in section~\ref{sec:THDM}. Furthermore we require that all charginos
and neutralinos are heavier than \unit{100}{GeV} and \unit{60}{GeV},
respectively. The light top squark is required to be heavier than
\unit{100}{GeV} while all other sfermions are taken to be heavier than \unit{350}{GeV}. Note that
bounds from fits to the oblique electroweak parameters are not applicable here,
because the MSSM contains new particles which couple directly to SM fermions.

Within these constraints we obtain the following
  upper limits on the  cross
sections  of the LHC reactions  $pp\to A \to VV'$:
\begin{align}\label{eq:mssm:A0limits}
    \sigma(pp\to A\to WW) &\lesssim \unit{2.0}{fb}\eqpunct,
  & \sigma(pp\to A\to ZZ) &\lesssim \unit{0.33}{fb}\eqpunct,
  \nonumber\\
    \sigma(pp\to A\to\gamma\gamma) &\lesssim \unit{0.27}{fb}\eqpunct,
  & \sigma(pp\to A\to Z\gamma) &\lesssim \unit{0.75}{fb}\eqpunct.
\end{align}
As mentioned earlier in this section, the maximal values are obtained for
$\tan\beta\sim 20$ and values slightly above this number. In accordance with \cite{Gunion:1991cw} we find that the
 non-SUSY contributions to the $A\to VV'$ decays lead to branching ratios below
$10^{-6}$ in this region of parameter space. However, the contributions from
loops of charginos and neutralinos can increase the branching ratios to the
order of $10^{-5}$. The largest $A\to VV'$ partial decay widths are obtained if $m_A$
is close to a two-chargino \emph{and} a two-neutralino threshold, while the
largest $A$ production rates are obtained for small $m_A$. Therefore, parameter
space regions with the largest $pp\to A\to VV'$ cross sections are characterized
by
\begin{equation}
  m_{\chi^\pm_1} \approx m_{\chi^0_2} \approx \unit{100}{GeV}
  \quad,\quad
  m_A \approx \unit{200}{GeV}
  \eqpunct.
\end{equation}
At $\tan\beta=20$ this is realized for
\begin{equation}
  M_2 \approx \unit{127}{GeV}
  \quad,\quad
  \mu \approx \unit{220}{GeV}
  \eqpunct.
\end{equation}
The other parameters from \eqref{eq:mssm:parameters} have only a weak influence
on  $\sigma_{A\to VV'}$, but must of course be chosen appropriately to
satisfy the experimental bounds on the pMSSM parameter space. 
  In the subset of this space we
consider here, the phenomenologically acceptable
   models typically contain a light top squark with
$m_{{\tilde t}_1} \sim \unit{120}{GeV}$ and a heavy stop with
   a mass between \unit{550}{GeV} and \unit{600}{GeV}.
%
%
%
\section{Top-colour assisted technicolour}
 \label{sec:TC2}
%
%
An alternative to the Higgs mechanism is EWSB triggered by the condensation of
(new) fermion-antifermion pairs.  Phenomenologically viable scenarios of this type
include models
  based on the concept of top-colour assisted
  technicolour\footnote{Again, to avoid 
  misunderstandings: we consider here
  TC2 models with three quark-generations.}   (TC2)
\cite{Hill:1994hp},  \cite{Cvetic:1997eb,Hill:2002ap}. These models
have two separate strongly interacting
sectors in order to explain EWSB and the large top-quark mass. Technicolour
interactions (TC) are responsible for most of EWSB via the condensation of
techni-fermions, $\langle {\bar T} T\rangle$ ($T = U, D$), but contribute very
little to the top-quark mass $m_t$. The top-colour interactions generate the bulk
of $m_t$ through condensation of top-quark pairs $\langle {\bar t} t \rangle$,
but make only a small contribution to EWSB.  The spin-zero states of TC2
are bound-states of of $t, b$ and of the techni-fermions. These two sets of
bound-states form two $SU(2)_L$ doublets $\Phi_{TC}, \Phi _t$, whose couplings
to the electroweak gauge bosons and to $t$ are formally equivalent to those of
a two-Higgs doublet model. The physical spin-zero states include
\begin{itemize}
\item a heavy neutral scalar $H_{TC}$ with a mass of order 1 TeV,
\item a neutral scalar $H_t$ which is a ${\bar t} t$ bound state. Its mass is
  expected to be of the order $2 m_t$ when estimated \`a la
  Nambu-Jona-Lasinio, but could in fact be lighter \cite{Chivukula:1998wd}.
\item a neutral ``top-pion'' $\Pi^0$ and a pair of charged ones, $\Pi^\pm$,
  whose masses are predicted to be of the order of a few hundred GeV
  \cite{Hill:1994hp,Buchalla:1995dp}.
\end{itemize}
 Several variants of TC2 were discussed in the literature,
 \cite{Hill:2002ap,Hill:1994hp}. Below we consider for definiteness
 TC2 with one family of technifermions.

The couplings of spin-zero states to the weak gauge bosons and to the $t$ and
$b$ quarks can be obtained from an effective ${\rm SU(2)_L \ \times U(1)_Y}$
invariant Lagrangian involving the doublets $\Phi_{TC}, \Phi _t$
\cite{Leibovich:2001ev}.  The interactions of the top quark with $H_t$ and
$\Pi^0$ are given by:
\begin{equation} \label{yuk-tc2}
  \Lcal_{Y,t} = 
  - \frac{Y_t}{\sqrt 2} {\bar t} t \,  H_t
  -  \frac{Y_\pi}{\sqrt 2}  {\bar t}i\gamma_5 t \,\Pi^0
  \eqpunct,
\end{equation}
where $Y_\pi =(Y_t v_T -\epsilon_t f_\pi)/v$ and $(Y_t f_\pi + \epsilon_t
v_T)/\sqrt{2} = m_t.$ Here $f_\pi$ denotes the value of the top-quark condensate
which is estimated in the TC2 models to lie between $40 \, {\rm GeV} \lesssim
f_\pi \lesssim 80 \, {\rm GeV}$ \cite{Hill:1994hp,Leibovich:2001ev}. Once
$f_\pi$ is fixed, $v_T$ is determined by the EWSB requirement that $f_\pi^2 +
v_T^2 = v^2 = (\unit{246}{GeV})^2$. The parameter $\epsilon_t$ denotes the
technicolour contribution to the top mass which is small by construction. The
large top-quark mass thus amounts to large top Yukawa couplings $Y_t$, $Y_\pi$,
e.g., $Y_t\approx Y_\pi\approx 3$ for $f_\pi\simeq$ 70 GeV and small
$\epsilon_t$.  On the other hand the
couplings of $H_t$ and $\Pi^0$ to $b$ quarks are significantly suppressed as
compared with the SM Higgs $b\bar b$ coupling. By construction, the top-colour
interactions do not generate a direct contribution to the mass of the
$b$ quark. In  TC2 models, the mass  of the $b$ quark is  due to extended
technicolour interactions and to  top-colour instanton effects.
 One may use  the following effective coupling of $\Pi^0$ to $b$ quarks:
\begin{equation}
  \Lcal_{Y,b} = - \epsilon_b \frac{f_\pi}{\sqrt{2}v} {\bar b}i
    \gamma_5 b \, \Pi^0
  \eqpunct,
\end{equation}
where $\epsilon_b = m_b \sqrt{2}/v_T$. With $m_b=\unit{4.8}{GeV}$ and
$f_\pi\leq\unit{80}{GeV}$ one gets $\epsilon_b \leq 0.03$.

Experimental constraints on the TC2 models were analyzed in 
 \cite{Burdman:1997pf,Yue:2000ay, Balaji:1995rh, Wu:1994dj}. 
The relevant constraints come from $b\to s\gamma$
decays, the LEP measurement of the hadronic $Z\to b\bar b$ branching
  ratio $R_b$
and the oblique electroweak parameter $T$. 
The bound from $b\to s\gamma$ decays is satisfied if
$\epsilon_t\lesssim 0.1$ \cite{Balaji:1995rh}. The bounds on the
   parameters of the TC2 models  that result from $R_b$ and  $T$
 are considerably weaker than the corresponding ones in the 2HDM due to
  additional contributions from extended technicolour 
  and topcolour gauge bosons
\cite{Yue:2000ay, Wu:1994dj}. In \cite{Yue:2000ay} it was found that top-pion
masses as low as \unit{280}{GeV} are still allowed for $\epsilon_t=0.1$ and
$f_\pi=\unit{70}{GeV}$. In order to estimate the maximal values of the LHC  $\Pi^0$ and $H_t$
production cross sections\footnote{In \cite{Belyaev:2005ct} the
  hadronic production of light techni-pions and their decays, in
  particular to two photons, were investigated within several
  technicolour models.},   we have therefore chosen these
  parameter values. 

\begin{figure}
  \centering
  \includegraphics{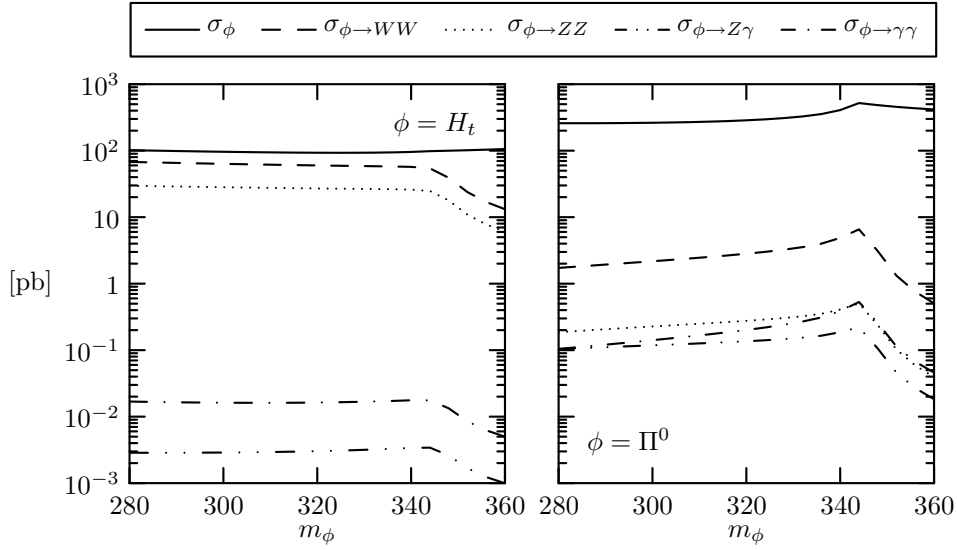}
  \caption{The LHC cross sections $\sigma(pp\to\phi)$ and $\sigma(pp\to\phi\to
    VV')$ as functions of $m_\phi$ (with $\phi\in\{H_t,\Pi^0\}$ and
    $VV'\in\{WW,ZZ,\gamma\gamma,Z\gamma\}$) in  TC2  with
    $\epsilon_t=0.1$ and $f_\pi=\unit{70}{GeV}$.}
  \label{fig:tc2:mA0}
\end{figure}

The main partonic production reaction of the neutral top-pion and of
$H_t$ is gluon fusion.
 In many TC2 models \cite{Hill:1994hp,Hill:2002ap}, technifermions do
  not have QCD charges and do therefore not contribute to this reaction.
  The amplitudes $gg \to \Pi^0, \ H_t$ are then  dominated by  top-quark loops 
  \cite{Hill:1994hp,Burdman:1999sr,Hashimoto:2002cy,Cao:2002af}. 
As  to $gg \to  H_t$, the contribution of
  topcolour gauge bosons is negligible \cite{Hashimoto:2002cy}. 

On the other hand, technifermions do contribute to the decays of the
top-pion $\Pi^0$ into $\gamma \gamma, \, Z\gamma, \, ZZ$. The
technicolour component of the mass eigenstate  $\Pi^0$, which is part
of a $SU(2)_L$ triplet, has effective couplings
   to weak gauge bosons through the chiral anomaly.  These
 ``anomalous'' terms depend on the specific technifermion sector of TC2, and
  can be determined by the respective chiral anomaly of the associated
  currents \cite{Dimopoulos:1980yf,Ellis:1980hz}. Here we consider for
  definiteness one family of technifermions. The respective anomalous
  contributions  to $\Pi^0 \to \gamma \gamma, \, Z\gamma, \, ZZ$ are
  readily computed; they can be found, for instance, in \cite{Ellis:1980hz,Chivukula:1995dt}.
 (Notice that  the anomaly factor for $\Pi^0 \to W^+W^-$ is zero.)

 Figure~\ref{fig:tc2:mA0} shows the LHC production cross sections for $H_t$,
$\Pi^0$ and the cross section times branching ratios for the $WW$, $ZZ$,
$\gamma\gamma$ and $Z\gamma$ final states as functions of the mass of the
 respective resonance  for the parameters given above. We see that the $H_t$ and
$\Pi^0$ production rates are quite large with values of \unit{100}{pb} and 200
to \unit{500}{pb}, respectively. If $m_{H_t}$ is below the $t\bar t$ threshold,
the $H_t$ decays dominantly into $WW$ and $ZZ$ for the above parameter
values, and we obtain cross sections
$\sigma(pp\to H_t\to WW,ZZ)$ of about \unit{60}{pb} and \unit{25}{pb},
respectively. The loop-mediated decays into $\gamma\gamma$ and $Z\gamma$ have
only very small branching ratios. The top-pion $\Pi^0$ decays dominantly into
gluon pairs if $\Pi^0\to t\bar t$ is kinematically forbidden. The cross
sections become maximal if the mass of the decaying particle is just below the
$t\bar t$ threshold. For $m_{\Pi^0}=m_{H_t}=\unit{340}{GeV}$ we have
\begin{align}\label{eq:tc2:pi0xsect}
    \sigma(pp\to \Pi^0\to WW) &= \unit{4.9}{pb}\eqpunct,
 & \sigma(pp\to H_t\to WW) &= \unit{57}{pb}\eqpunct,
  \nonumber\\
   \sigma(pp\to \Pi^0\to ZZ) &= \unit{0.41}{pb}\eqpunct,
 & \sigma(pp\to H_t\to ZZ) &= \unit{26}{pb}\eqpunct,
  \nonumber\\
    \sigma(pp\to \Pi^0\to\gamma\gamma) &= \unit{0.39}{pb}\eqpunct,
  & \sigma(pp\to H_t\to\gamma\gamma) &= \unit{0.02}{pb}\eqpunct,
 \nonumber \\
   \sigma(pp\to \Pi^0\to Z\gamma) &= \unit{0.19}{pb}\eqpunct,
 & \sigma(pp\to H_t\to Z\gamma) &= \unit{0.003}{pb}\eqpunct.
 \end{align}
 Without the anomalous contributions  $\sigma_{\Pi^0 \to \gamma \gamma}$
  would be larger by a factor $\sim 2.5$, while  $\sigma_{\Pi^0 \to ZZ}$
 would decrease by about a factor  $\sim 2$. 
 The cross section $\sigma_{ \Pi^0 \to Z\gamma}$ remains essentially
 unchanged.

 It is worth emphasizing that the cross sections
  for  $\Pi^0\to \gamma \gamma, \, Z\gamma$ are much 
   larger than the corresponding ones for $H_t$. 
    As to the possible
  size of the ratio  $\sigma_{\Pi^0\to VV}/\sigma_{H_t \to VV}$ 
   ($V=W,Z$) the
   values in (\ref{eq:tc2:pi0xsect}) are, for the specific
   technifermion  sector,  rather conservative. This
   ratio increases for  smaller values of $f_\pi$.

If $\Pi^0$ or $H_t$ are heavier than $2m_t$ they decay dominantly into $t\bar t$
pairs and the branching ratios of the  other decay modes become very small.
%
%
%
\section{Summary and Conclusions} \label{sec:conclusions}
%
%
We have computed  and analyzed the LHC cross sections for 
 the production of a heavy 
pseudoscalar Higgs boson  $A$, and also those of a heavy scalar $H$,
   and their subsequent decays into electroweak gauge
bosons in several SM extensions. We determined and scanned the phenomenologically
   allowed regions of the corresponding parameter spaces in order to find the largest possible
values of these cross sections. Within the non-SUSY models analyzed
here  we considered spin-zero states $A$, $H$
   with masses $m_{A,H} \lesssim 2 m_t$, for reasons stated in Section~\ref{sec:approx}.

For models with elementary Higgs fields, the  largest cross sections $\sigma_{A\to VV'}$ 
  for $A$ were found in the decoupling limit of a
2HDM with a fourth generation of chiral fermions and $\tan\beta\approx 6.3$. The
total LHC production rate for $A$ can become of the order  of  \unit{800}{pb}
in this scenario, without violating bounds from direct Higgs-boson searches at the
Tevatron. The signal cross sections for the decays of $A$ into electroweak gauge bosons
can then be of the order of a few picobarn.

In the 3-generation type-II 2HDM we found the 
  largest  cross sections $\sigma_{A\to VV'}$ in the
decoupling limit with $\tan\beta\approx 0.75$. For the $WW$ and $\gamma\gamma$
final states the maximum values are of the order of \unit{0.1}{pb}. For the $ZZ$
and $Z\gamma$ final states they are smaller by about one order of magnitude.
Extending the 2HDM by vector-like quarks in the way as was  done in
   \cite{Bernreuther:2009ts} does not change the maximum allowed
values of the cross sections significantly.

In the MSSM we obtained the largest pseudoscalar cross sections for large
$\tan\beta$ and for a spectrum where $m_A$ is close to both a two-chargino and a
two-neutralino threshold. Loops of charginos and neutralinos then yield the
dominant contributions to the $A\to VV'$ decay rates. However, the resulting
   signal cross sections  are at most a few femtobarn.

Finally we studied topcolour-assisted technicolour as a paradigm for models with
composite spin-zero states. Though conceptually very different, the ``Higgs sector''
of this model and the couplings of the spin-zero particles to top quarks
 and electroweak gauge bosons
   corresponds to a 2HDM with small $\tan\beta$. Experimental constraints on 
  the parameter space of these models from
measurements of $R_b$ and the $\rho$ parameter are relaxed due to contributions
from technicolour and topcolour gauge bosons. In a TC2 model with one
  family of technifermions, the maximal allowed signal cross
sections for decays of the pseudoscalar top-pion $\Pi^0$ into electroweak gauge
bosons are of the order of \unit{5}{pb} for $\Pi^0\to WW$
  and between  \unit{0.1}{pb} and  \unit{0.5}{pb} for 
$\Pi^0 \to Z\gamma, \, \gamma\gamma, \, Z Z$.  The
amplitude of the strongest  decay mode, $\Pi^0\to WW$, is actually
insensitive to the specific technifermion sector.

In conclusion we found that, with the exception of the MSSM, all the
models analyzed above  permit the  cross sections
for $pp \to A\to VV'$ to be of observable size at the LHC, in
particular for $A\to WW, \gamma \gamma$. The  cross sections $\sigma_{A\to
  \gamma\gamma}$  and $\sigma_{A\to Z \gamma}$ are typically one to
two orders of magnitude larger than the corresponding ones for a heavy
scalar $H$. Moreover, it is also possible  that $\sigma_{A\to
  WW}$  is of the same order of magnitude as  $\sigma_{H\to WW}$, as
  Figs.~\ref{fig:scatter.thdm} and~\ref{fig:scatter.thdm4} show.
  Obviously this does not mean that these are the most probable
     channels for discovering $A$.  Very likely, the discovery 
 modes would be $A\to  b{\bar b}, \ \tau^+\tau^- , \ Zh$, or $A\to {t \bar
   t}$,  depending on the mass spectra and coupling strengths.
  But a pseudoscalar resonance  would then be observable at 
  the LHC  also in  its decays into electroweak gauge bosons, in
  particular in $WW$ and  $\gamma \gamma$ final states.

%
\subsubsection*{Acknowledgments}

We would like to thank Alexander Belyaev, Jens Erler, Ulrich Haisch, J\"urgen Rohrwild,
 Oscar St{\aa}l, and Peter Zerwas for fruitful
discussions and information about their work. A special thanks goes to Karina Williams for patiently answering
questions about \code{FeynHiggs} and \code{HiggsBounds}. \\
This work was supported by Deutsche Forschungsgemeinschaft DFG SFB/TR9 and BMBF.
 P.G. is supported by a stipend from the DFG funded RWTH Graduiertenkolleg
 ``Elementarteilchenphysik an der TeV Skala''.

\bibliography{Apdecay}
\end{document}